\documentclass[aps,pra,reprint,superscriptaddress,longbibliography]{revtex4-1}
\usepackage{graphicx,color}
\usepackage{amsmath, amssymb}
\usepackage{longtable}
\usepackage{natbib}
\usepackage{bm}
\usepackage{hyperref}

\allowdisplaybreaks[1]

\begin{document}
\title{
Quadratic Jahn-Teller effect of fullerene anions
}
\author{Dan Liu}
\affiliation{Theory of Nanomaterials Group, University of Leuven, Celestijnenlaan 200F, B-3001 Leuven, Belgium}
\author{Yasuyuki Niwa}
\altaffiliation[Present address: ]{Institute for Chemical Research, Kyoto University, Uji, Kyoto 611-0011, Japan}
\affiliation{Undergraduate School of Industrial Chemistry, Faculty of Engineering, Kyoto University, Kyoto 606-8501, Japan}
\author{Naoya Iwahara}
\email{naoya.iwahara@gmail.com}
\affiliation{Theory of Nanomaterials Group, University of Leuven, Celestijnenlaan 200F, B-3001 Leuven, Belgium}
\author{Tohru Sato}
\affiliation{Fukui Institute for Fundamental Chemistry, Kyoto University, Takano-Nishihiraki-cho 34-4, Sakyo-ku, Kyoto 606-8103, Japan}
\affiliation{Department of Molecular Engineering, Graduate School of Engineering, Kyoto University, Kyoto 615-8510, Japan}
\affiliation{Unit of Elements Strategy Initiative for Catalysts \& Batteries, Kyoto University, Katsura, Nishikyo-ku, Kyoto 615-8510, Japan} 
\author{Liviu F. Chibotaru}
\email{liviu.chibotaru@gmail.com}
\affiliation{Theory of Nanomaterials Group, University of Leuven, Celestijnenlaan 200F, B-3001 Leuven, Belgium}
\date{\today}

\begin{abstract}
The quadratic Jahn-Teller effect of C$_{60}^{n-}$ ($n=$ 1-5) is investigated from the first principles. 
Employing the density functional theory calculations with hybrid functional, the quadratic vibronic coupling constants of C$_{60}^-$ were derived. 
The warping of the adiabatic potential energy surface of C$_{60}^-$ by the quadratic vibronic coupling is estimated 
to be 
about 2 meV, which is much smaller than the Jahn-Teller stabilization energy ($\approx$ 50 meV).
Because of the selection rule and the vibronic reduction, the quadratic coupling slightly modifies the vibronic states of C$_{60}$ anions.
Particularly, in the case of C$_{60}^{3-}$, parity and symmetry selection rule significantly reduces the effect of quadratic coupling on vibronic states. 
The present results confirm that the low-energy vibronic dynamics of C$_{60}^{n-}$ is of pseudorotational type. 
\end{abstract}

\maketitle

\section{Introduction}
\label{Introduction}
The Jahn-Teller (JT) effect \cite{Jahn1937, Bersuker1989, Chancey1997} plays a central role in the low-energy structures of fullerene C$_{60}$ anions and the electronic properties of fullerene based compounds. 
The relative importance of static and dynamical JT effect has been intensively investigated in molecular spectroscopy \cite{Kato1993, Kondo1995, Gunnarsson1995, Tomita2005, Tomita2006, Hands2008, Iwahara2010, Kern2013, Stochkel2013, Huang2014, Kundu2015},
in the superconducting and insulating alkali-doped fullerides \cite{Gunnarsson1997, Gunnarsson2004, Auerbach1994, Manini1994, OBrien1996, Winter1996, Brouet2001, Brouet2002c, Chibotaru2005, Wachowiack2005, Klupp2012, Iwahara2013, Potocnik2014, Iwahara2015, Zadik2015, Iwahara2016, Nomura2016, Nava2018, Matsuda2018},
in the ferromagnetic TDAE-C$_{60}$ \cite{Kawamoto1997, Sato1997}, where TDAE stands for tetrakis(dimethylamino)ethylene,
and other organic fullerene compounds \cite{Amsharov2011, Francis2012, Konarev2013}.
The dynamical $t_{1u}^n \otimes h_g$ JT model for C$_{60}^{n-}$ ions has been analyzed with various approaches. 
Within a linear vibronic model, the vibronic dynamics is of pseudorotational type \cite{Auerbach1994, OBrien1996}, while, by turning on the quadratic vibronic coupling, it becomes of tunneling splitting type \cite{Dunn1995, Sindi2007} due to the hindering of pseudorotation.

In order to assess the nature of JT effect in C$_{60}$ anions, the knowledge of vibronic coupling constants is essential. 
After many experimental \cite{Gunnarsson1995, Winter1996, Hands2008} and theoretical \cite{Varma1991, Schluter1992, Faulhaber1993, Antropov1993, Breda1998, Manini2001, Saito2002, Frederiksen2008, Janssen2010} studies, the linear orbital vibronic coupling parameters of C$_{60}^-$ were finally established recently \cite{Iwahara2010} 
(for details on this problem, see Refs. \cite{Iwahara2010, Gunnarsson2012}). 
In this work, the linear vibronic coupling parameters were derived by simulating high-resolution photoelectron spectrum \cite{Wang2005},
and the obtained parameters were found to agree well with the density functional theory (DFT) calculations using hybrid functional. 
Subsequent calculations within the {\it GW} approximation gave close values of stabilization energy for C$_{60}^-$ \cite{Faber2011}. 
On the other hand, the knowledge of nonlinear vibronic coupling parameters of C$_{60}$ anions is still lacking. 
Because of intermediate linear vibronic coupling in C$_{60}^-$ \cite{Iwahara2010}, the quadratic coupling is expected to be weak \cite{Iwahara2013}. 
Indeed, the photoelectron spectra of C$_{60}^{-}$ \cite{Gunnarsson1995, Iwahara2010} and infrared (IR) spectra of Mott-insulating Cs$_3$C$_{60}$ \cite{Matsuda2018} has been well reproduced without quadratic JT coupling.
On the other hand, near IR spectra \cite{Tomita2005, Tomita2006, Hands2008} of C$_{60}$ anions 
and electron paramagnetic resonance measurements of C$_{60}^-$ in solution \cite{Kundu2015} have been interpreted on the basis of tunneling splitting character of the low-lying vibronic spectrum.

The purpose of this work is to reveal the nature of the JT dynamics of C$_{60}^{n-}$ including both linear and quadratic vibronic coupling.
The quadratic vibronic coupling constants of C$_{60}$ anions were derived from the DFT calculations with B3LYP functional. 
With the obtained constants, the warping of the adiabatic potential energy surface of C$_{60}^-$ 
was
found to amount to a few meV, which is much smaller than the JT stabilization energy. 
It 
was also 
found that, because of the selection rule and the vibronic reduction, the quadratic coupling does not modify the vibronic spectrum of linear $t_{1u}^n \otimes 8h_g$ JT model.
The present results show that the quadratic vibronic coupling in C$_{60}$ anions is sufficiently weak for the JT dynamics to remain of pseudorotational type.

\section{Quadratic Jahn-Teller model for C$_{60}$ anions}
\subsection{bielectronic and vibronic interactions}
\label{Sec:H}
The neutral fullerene C$_{60}$ ($I_h$ symmetry) has triply degenerate $t_{1u}$ lowest unoccupied molecular orbitals (LUMO).
The LUMO levels are well separated from the other molecular orbital levels, therefore, the $t_{1u}$ shell model describes adequately the low-energy electronic structure of C$_{60}^{n-}$ anions ($n =$ 1-5).
The model Hamiltonian for C$_{60}^{n-}$ ion consists of 
the
bielectronic term $\hat{H}_\text{bi}$ and the vibronic term.
The former causes the term splitting of the $t_{1u}^n$ configurations.
The latter describes the vibronic coupling between the $t_{1u}$ orbitals and the nuclear vibrations of $a_g$ or $h_g$ type \cite{Jahn1937}.
Using the equilibrium structure of neutral C$_{60}$ as a reference, the vibronic term is represented as 
the
sum of following contributions \cite{Auerbach1994, Manini1994, Dunn1995, OBrien1996, Chancey1997}:
\begin{eqnarray}
 \hat{H}_\text{vib} &=& \sum_{\mu \Gamma \gamma} \frac{1}{2} \left(\hat{p}_{\Gamma(\mu) \gamma}^2 + K^{\Gamma}_{\mu} \hat{q}_{\Gamma(\mu) \gamma}^2 \right),
\label{Eq:Hvib}
\\
 \hat{H}_\text{JT}^{(1)} &=& \sum_{\lambda \lambda' \sigma} \sum_{\mu \gamma} {V}^{h_g}_{\mu}
                        \hat{c}_{\lambda \sigma}^\dagger \hat{c}_{\lambda' \sigma} \hat{q}_{h_g(\mu)\gamma}
                        \sqrt{\frac{5}{2}} \langle t_{1u}\lambda'| h_g \gamma  t_{1u}\lambda \rangle, 
\nonumber\\
\label{Eq:HLJT}
\\
 \hat{H}_\text{JT}^{(2)} &=& \sum_{\lambda \lambda' \sigma} 
                      \sum_{\mu_i \Gamma_i} 
                      \sum_{\nu \gamma} {V}^{\Gamma_1 \Gamma_2}_{\nu \mu_1 \mu_2}
                      \hat{c}_{\lambda \sigma}^\dagger \hat{c}_{\lambda' \sigma} 
\nonumber\\
                    &\times&
                  \{\hat{q}_{\Gamma_1(\mu_1)} \otimes \hat{q}_{\Gamma_2(\mu_2)} \}_{\nu h_g \gamma}
                        \sqrt{\frac{5}{2}} \langle t_{1u}\lambda'| h_g \gamma  t_{1u}\lambda \rangle. 
\label{Eq:HQJT}
\end{eqnarray}
Here, $\lambda = x,y,z$ and $\gamma = \theta, \epsilon, \xi, \eta, \zeta$ are the components of the $t_{1u}$ and $h_g$ irreducible representations of $I_h$ group, respectively, 
$\hat{q}$ and $\hat{p}$ are the mass-weighted normal coordinates and momenta of the reference system, respectively, $\mu$ denotes the vibrational modes of the same symmetry,
$\{\hat{q}_{\Gamma_1} \otimes \hat{q}_{\Gamma_2}\}_{\nu h_g\gamma}$ are symmetrized products (see for explicit form Ref. \cite{SM}),
$\langle t_{1u}\lambda'| h_g \gamma  t_{1u}\lambda \rangle$ are the Clebsch-Gordan coefficients, 
and $K^\Gamma$, $V^\Gamma$ and $V^{\Gamma_1 \Gamma_2}$ are the force constant, linear and quadratic vibronic coupling parameters, respectively.
The coefficient $\sqrt{5/2}$ is introduced in vibronic terms following Refs. \cite{OBrien1969, OBrien1971, Auerbach1994, Manini1994, OBrien1996, Chancey1997}, which results in $\sqrt{2/5}$ times smaller vibronic coupling parameters than those in Ref. \cite{Dunn1995}.
As the basis of 
the
$h_g$ representation, atomic $d$ functions are used \cite{OBrien1969, OBrien1971, Auerbach1994, Manini1994, OBrien1996, Chancey1997}: $\theta$, $\epsilon$, $\xi$, $\eta$, $\zeta$ components of the $h_g$ representation transform as $(2z^2-x^2-y^2)/\sqrt{6}$, $(x^2-y^2)/\sqrt{2}$, $\sqrt{2} yz$, $\sqrt{2} zx$, $\sqrt{2} xy$ components of the $d$ orbitals, respectively, under symmetric operations of $I_h$ group (The Cartesian coordinate axes correspond to the $C_2$ axes of C$_{60}$ as in Refs. \cite{Boyle1980, Altmann1994, Gunnarsson2004}).
Eq. (\ref{Eq:HQJT}) contains two parameters \cite{Dunn1995} since the symmetric square of $h_g$ contains two $h_g$ ($\nu = 1,2$) \cite{Altmann1994}. 
The symmetrized polynomial $\{\hat{q}_{h_g(\mu_1)} \otimes \hat{q}_{h_g(\mu_2)} \}_{1 h_g \gamma}$ and $\{\hat{q}_{h_g(\mu_1)} \otimes \hat{q}_{h_g(\mu_2)} \}_{2 h_g \gamma}$ become zero for the deformation along the $D_{3d}$ and $D_{5d}$ minima, respectively, as in Ref. \cite{Dunn1995}. 
The vibronic Hamiltonian for the unimportant totally symmetric modes is not written here for simplicity. 

It is convenient to write down the vibronic Hamiltonian in the basis of electronic terms, particularly, when there is term splitting \cite{OBrien1996, Chancey1997}. 
The orbital part for C$_{60}^{n-}$ can be written as follows:
\begin{eqnarray}
 \hat{H}_\text{JT}^{(1)} &=& 
                      \sum_{\mu \gamma} V^{h_g}_{\mu}
                      \hat{q}_{h_g(\mu)\gamma} \hat{C}^{(n)}_{\gamma},
\label{Eq:HLJTmat}
\\
 \hat{H}_\text{JT}^{(2)} &=& 
                      \sum_{\mu_i \Gamma_i} 
                      \sum_{\nu \gamma} V^{\Gamma_1 \Gamma_2}_{\nu \mu_1 \mu_2}
                      \{\hat{q}_{\Gamma_1(\mu_1)} \otimes \hat{q}_{\Gamma_2(\mu_2)} \}_{\nu h_g \gamma} \hat{C}^{(n)}_{\gamma}.
\label{Eq:HQJTmat}
\end{eqnarray}
where, $\hat{C}_{\gamma}^{(n)}$ are the matrices of 
the
Clebsch-Gordan coefficients, which are given in Refs. \cite{OBrien1996, Chancey1997} and also in Supplemental Materials \cite{SM}.
For the derivation of the matrices in Eqs. (\ref{Eq:HLJTmat}) and (\ref{Eq:HQJTmat}) in the basis of $t_{1u}^n$ terms, see Ref. \cite{Liu2018}.

\subsection{Adiabatic potential energy surface} 
\label{Sec:APES}
The structure of the adiabatic potential energy surface (APES) provides fundamental information about the nature of vibronic dynamics. 
According to Liehr's minimax rule \cite{Liehr1963I}, the JT distortion which lifts the degeneracy of the initial electronic state corresponds to the one maintaining the highest subgroup of the system.
In the case of $I_h$ system with $T_{1u}$ electronic state ($n = 1,5$), the corresponding JT deformation is accordingly either of $D_{5d}$ or of $D_{3d}$ type.
Further symmetry lowering may arise due to the higher order vibronic coupling or pseudo Jahn-Teller coupling \cite{Muramatsu1970}. 
The effect of the warping of the APES can be modeled using the invariants of $I_h$ group, which indeed shows the symmetry of the JT deformed structure either of $D_{5d}$ or of $D_{3d}$ \cite{Chancey1997}.
This was unambiguously proved by explicitly analyzing the quadratic JT Hamiltonian \cite{Dunn1995}. 
The first ($\nu = 1$) and the second ($\nu = 2$) quadratic vibronic couplings in Eq. (\ref{Eq:HQJT}) favors the $D_{5d}$ and $D_{3d}$ deformations, respectively. 
If the energy difference between them is large, the system tends to be localized at the potential's minima. 

So far, the analysis of the APES has been done only taking into account the JT active $h_g$ mode \cite{Chancey1997, Dunn1995, Alqannas2013, Dunn2015}. 
However, in the case of $I_h$ system, the JT inactive modes can be included in Eq. (\ref{Eq:HQJT}). 
This is also understood based on subduction of 
irrep.
Under $I_h\downarrow D_{5d}$, only the $h_g$ representation contains the totally symmetric representation, whereas under 
$I_h\downarrow D_{3d}$ or $I_h\downarrow D_{2h}$,
totally symmetric representation also appears from the $g_g$ representation \cite{Altmann1994}. 
Below, the effect of the $g_g$ mode on the APES is analyzed within $t_{1u} \otimes (g_g \oplus h_g)$ JT model. 
For simplicity, the index $g$ for parity is omitted in the coupling parameters.
Since the $g_g$ mode does not contribute to the $D_{5d}$ deformation, the energy and the JT deformation at the $D_{5d}$ minima are the same as those for $t_{1u} \otimes h_g$ JT model \cite{Dunn1995}:
\begin{eqnarray}
 U_{D_{5d}} &=& - \frac{\left(V^{h}\right)^2}{2K^{h} - \frac{8}{\sqrt{5}} V^{hh}_1},
\label{Eq:UD5d}
\\
 q_{h_g} &=& \frac{V^{h}}{K^{h} - \frac{4}{\sqrt{5}} V^{hh}_1}, \quad q_{g_g} = 0.
\end{eqnarray}
On the other hand, the energy at 
the
$D_{3d}$ minima is modified by the $g_g$ contribution as
\begin{eqnarray}
 U_{D_{3d}} &=& -\frac{\left(V^{h}\right)^2}{2 \left(K^{h} - \frac{4}{3} V_2^{hh}\right) - \frac{8\left(V^{gh}_1\right)^2}{9 K^{g} + 15 V^{gg}} },
\label{Eq:UD3d}
\\
 q_{h_g} &=& \frac{V^{h}}{K^{h} - \frac{4}{3} V^{hh}_2 - \frac{4\left(V^{gh}_1\right)^2}{9 K^{g} + 15 V^{gg}}}, 
\nonumber\\
 q_{g_g} &=& \frac{6 V^{h} V^{gh}_1}{(3K^{g} + 5V^{gg}) (3K^{h} - 4V^{hh}_2) - 4\left(V^{gh}_1\right)^2}. 
\end{eqnarray}
The coupling to the $g_g$ mode lowers the $D_{3d}$ minima. 
In the case of the $D_{2h}$ minima, two components of $h_g$ and one of $g_g$ become totally symmetric. 
The analytical expression of the energy at 
the
$D_{2h}$ minima becomes cumbersome, thus, approximate forms 
of $U$ and $q$ 
including up to the second order terms in quadratic vibronic couplings are given here:
\begin{eqnarray}
 U_{D_{2h}} &=& -\frac{\left(V^{h}\right)^2}{2K^{h}}\left[ 1 + \frac{\sqrt{5} V^{hh}_1 + 9 V^{hh}_2}{8 K^{h}} 
 \right.
\nonumber\\
 &+& 
 \left.
  \frac{3 \left(V^{gh}_1\right)^2}{8 K^{g} K^{h}} + \frac{\left(V^{hh}_1\right)^2 + 3 \left(V^{hh}_2\right)^2}{2\left(K^{h}\right)^2} 
 \right], 
\label{Eq:UD2h}
\\
 q_{h_g\theta} &=& \frac{V^{h}}{K^{h}} \left[ 1 + \frac{\sqrt{5} V^{hh}_1 + 9 V^{hh}_2}{8 K^h} + \frac{3 \left(V^{gh}_1\right)^2}{K^g K^h} 
 \right.
\nonumber\\
 &+& 
 \left. 
 \frac{\left(V^{hh}_1\right)^2 + 3\left(V^{hh}_2\right)^2}{2\left(K^g\right)^2}
 \right],
\nonumber\\
 q_{h_g\epsilon} &=& \frac{\sqrt{3} V^h}{8 K^h} \left[\frac{3V^{hh}_1 - \sqrt{5} V^{hh}_2}{K^h} + \frac{\sqrt{5} V^{gh}_1 V^{gh}_2}{K^g K^h} \right],
\nonumber\\
 q_{g_g a} &=& \frac{1}{2}\sqrt{\frac{3}{2}} \frac{V^h}{K^h} \left[ \frac{V^{gh}_1}{K^g} + \frac{\left(\sqrt{5} V^{hh}_1 + 9 V^{hh}_2 \right) V^{gh}_1}{8K^g K^h} 
 \right.
\nonumber\\
 &+&
 \left.
  \frac{\left(3\sqrt{5} V^{hh}_1 - 5 V^{hh}_2 \right) V^{gh}_2}{8K^g K^h} \right].
\end{eqnarray}

The energy difference between the $D_{5d}$ and $D_{3d}$ minima is 
\begin{eqnarray}
 \Delta U &=& U_{D_{3d}} - U_{D_{5d}} 
\nonumber\\
 &=& \frac{\left(V^{h}\right)^2}{2K^{h}} 
 \left[
  \frac{6V^{hh}_1 - 2\sqrt{5} V^{hh}_2}{3\sqrt{5} K^{h}} 
 - 
  \frac{2 \left(V^{gh}_1\right)^2}{9K^{g} K^{h}} 
 \right.
\nonumber\\
 &+&
 \left.
   \frac{8 \left(V^{hh}_1\right)^2}{5\left(K^{h}\right)^2} 
 + \frac{5 \left(V^{hh}_2\right)^2}{9\left(K^{h}\right)^2}
 \right].
\label{Eq:DeltaU}
\end{eqnarray}
The terms up to the second order of quadratic couplings are taken into account.
Considering the usual situation where $|V^{\Gamma_1 \Gamma_2}| < |K^\Gamma|$, the effect of the $g_g$ mode on 
the
APES would be minor. 

In the case of $n =$ 2-4, the quadratic coupling becomes a few times stronger than that in $n = 1,5$. 
Due to the presence of the multiplet splitting, the structure of the APES becomes more complicated \cite{Alqannas2013}. 
The symmetry of the potential minima for $n = 2,4$ is $D_{5d}$, $D_{3d}$, or $D_{2h}$ within $t_{1u}^{2/4} \otimes h_g$ JT model. 
Therefore, 
the
JT inactive $g_g$ modes would play a similar role as for $n = 1,5$.
On the other hand, the potential minima of $t_{1u}^3 \otimes h_g$ JT model is either $D_{2h}$ or $C_{2h}$, and in the latter, $t_{1g}$ and $t_{2g}$ as well as $g_g$ vibrations can modify the APES. 
For $n = $ 2-4, we will not analyze the APES further.

\subsection{Vibronic states}
\label{Sec:Vibro}
The quadratic vibronic coupling modifies the vibronic states. 
As discussed in the previous section, the effect of the non-JT modes is expected to be weak, thus, only the JT active $h_g$ modes are taken into account here. 
The nature of the linear vibronic states of the dynamical JT Hamiltonian,
\begin{eqnarray}
 \hat{H}^{(0)} &=& \hat{H}_\text{bi} + \hat{H}_\text{vib} + \hat{H}_\text{JT}^{(1)},
\end{eqnarray}
has been intensively studied. 
Because of the SO(3) symmetry of the linear $T_{1u} \otimes h_g$ JT Hamiltonian \cite{OBrien1971, Pooler1980}, the Hamiltonian, one of the vibronic angular momenta, $\hat{J}_z$, and the square of the angular momentum $\bm{\hat{J}}^2 = \hat{J}_x^2 + \hat{J}_y^2 + \hat{J}_z^2$ are of commuting observables.
Besides, in the case of C$_{60}^{3-}$, the seniority of the electronic terms \cite{Racah1943} is inherited in the vibronic state and expressed by parity \cite{Iwahara2013, Liu2018}: 
\begin{eqnarray}
 \hat{P} &=& \left(\hat{I}_{T_{1u}} - \hat{I}_{H_u} \right) \prod_{\mu=1}^8 \prod_{\gamma = \theta, \epsilon, \xi, \eta, \zeta}  (-1)^{\hat{n}_{h_g(\mu)\gamma}}.
\label{Eq:P}
\end{eqnarray}
Here, $\hat{I}_\Gamma$ is the projection operator into $\Gamma$ electronic term ($\Gamma = T_{1u}, H_u$) and $\hat{n}_{h_g(\mu)\gamma}$ is the $h_g(\mu)\gamma$ vibrational quantum number operator.
Therefore, the linear vibronic states of C$_{60}^{n-}$ ($n = 1, 2, 4, 5$) are characterized by the magnitude of the angular momentum $J$ $(= 0, 1, 2, ...)$, its $z$ component $M_J$ $(= -J, -J + 1, ..., J)$, and the principal quantum number $\alpha$ which distinguishes the energy levels.
In the case of C$_{60}^{3-}$, parity $P$ $(= \pm 1)$ is added to the set of quantum numbers. 

The total angular momenta do not commute with the quadratic vibronic Hamiltonian, $\hat{H}_\text{JT}^{(2)}$. 
Thus, the degeneracy of linear vibronic states is partly lifted, and split vibronic states are characterized by the irrep of $I_h$ group. 
In the case of C$_{60}^{3-}$, the parity (\ref{Eq:P}) does not commute with $\hat{H}^{(2)}_\text{JT}$:
\begin{eqnarray}
 \hat{P} \hat{H}_\text{JT}^{(2)} \hat{P} = - \hat{H}_\text{JT}^{(2)}.
\label{Eq:PH2}
\end{eqnarray}
The vibronic states of the full Hamiltonian, 
\begin{eqnarray}
\hat{H} = \hat{H}^{(0)} + \hat{H}^{(2)}_\text{JT}, 
\end{eqnarray}
are expressed by the superposition of the linear vibronic states. 
To describe the vibronic states of $\hat{H}$, it is convenient to use the irreducible linear vibronic states, $|\Psi^{(0)}_{\alpha J \Gamma \gamma (P)}\rangle$, where $\Gamma$ is an irrep included in $J \downarrow I_h$, and $\gamma$ is a component of $\Gamma$. 
The vibronic states are written
\begin{eqnarray}
 |\Psi_{\alpha \Gamma \gamma}\rangle &=& \sum_{\beta J (P)} |\Psi^{(0)}_{\beta J \Gamma \gamma (P)} \rangle C_{\beta J \Gamma \gamma (P); \alpha}.
\label{Eq:Psi}
\end{eqnarray}
Coefficients $C_{\beta J\Gamma \gamma (P); \alpha}$ are determined from the diagonalization of the $\hat{H}$ matrix in the basis of $|\Psi^{(0)}_{\alpha J \Gamma \gamma (P)}\rangle$.
There are two selection rules for the matrix elements of $\hat{H}^{(2)}_\text{JT}$.
First, since $\hat{H}^{(2)}_\text{JT}$ is totally symmetric, the off-diagonal blocks between different irrep are zero. 
Second, in the case of C$_{60}^{3-}$, due to Eq. (\ref{Eq:PH2}), only the off-diagonal blocks between the $|\Psi^{(0)}_{\alpha J \Gamma \gamma P}\rangle$'s characterized by the opposite parities are nonzero. 
Accordingly, we have
\begin{eqnarray}
 \langle \Psi^{(0)}_{\beta J\Gamma \gamma (P)}| \hat{H}^{(2)}_\text{JT} | \Psi^{(0)}_{\beta' J' \Gamma' \gamma' (P')} \rangle 
 =
 \delta_{\Gamma \Gamma'} \delta_{\gamma \gamma} (\delta_{P,-P'}) 
\nonumber\\
\times
 \langle \Psi^{(0)}_{\beta J\Gamma \gamma (P)}| \hat{H}^{(2)}_\text{JT} | \Psi^{(0)}_{\beta' J'\Gamma \gamma (-P)} \rangle.
\label{Eq:mat}
\end{eqnarray}
Here, the expressions in the parentheses are only considered for C$_{60}^{3-}$.
Thus, the vibronic levels of $n = 1,2,4,5$ ($n = 3$) change linearly (quadratically) with the strength of 
the
quadratic vibronic coupling. 

\section{DFT calculations of quadratic vibronic coupling parameters}
\label{Sec:method}
In order to derive the quadratic vibronic coupling parameters, DFT method with hybrid B3LYP exchange correlation functional \cite{Becke1993} and triple-zeta basis set (6-311G(d)) was employed. 
All DFT calculations were done using {\tt Gaussian} 16 program \cite{g16}.
The linear orbital vibronic coupling parameters of C$_{60}^-$ calculated by the hybrid B3LYP functional are close to the experimental data \cite{Iwahara2010}.
The electron affinity derived with the Delta-SCF approach to C$_{60}^-$ and C$_{60}$ is 2.497 eV, which is in line with the recent experimental value 2.683 eV \cite{Wang2005, Huang2014}.

The quadratic vibronic coupling constants were derived within three approaches:
\begin{enumerate}
\item[(i).]   Fitting of the APES of deformed C$_{60}^-$ to model JT Hamiltonian, 
\item[(ii).]  Fitting of the $t_{1u}$ LUMO level of deformed neutral C$_{60}$ to model JT Hamiltonian,
\item[(iii).] Comparing the Hessian for 
the
relaxed C$_{60}^-$ structure and for the model JT Hamiltonian. 
\end{enumerate}
In the first approach, the APES of deformed C$_{60}^-$ along various modes are fitted to the quadratic JT Hamiltonian. 
The deformations are both along single $a_g/h_g$ mode and also linear combinations of two modes ($a_g$ and $h_g$ or two different $h_g$).
This approach uses the symmetry broken electronic wave functions, which allows to take into account the electron correlation, however, artificial error might be introduced.
The linear vibronic coupling parameters of C$_{60}^-$ were derived by a similar method using the gradient of 
the
APES at high-symmetric structure in Ref. \cite{Iwahara2010}. 
The second method is similar to the first one but with the use of 
the
$t_{1u}$ LUMO of neutral C$_{60}$. 
The advantage of the method is that the obtained parameters are free from the artificial error owing to the symmetry breaking of the wave function, whereas, the effect of the relaxation of the molecular orbitals by electron doping is neglected. 
As seen in the previous studies on 
the
linear vibronic coupling parameters, the coupling parameters from method (i) \cite{Iwahara2010} and method (ii) \cite{Saito2002, Janssen2010} do not differ much in prediction.
In the last approach, the quadratic coupling parameters are derived from the Hessian at 
the
JT deformed structure, since the Hessian at the equilibrium high-symmetric structures 
contains
not only $K^\Gamma$ but also $V^{\Gamma_1 \Gamma_2}$ (Appendix \ref{A:Hessian}). 
By comparing the expression of 
the
second derivatives within model and the DFT-derived Hessian at the corresponding potential minima, the parameters can be determined. 
In this work, the Hessians at 
the
$D_{2h}$ and $D_{3d}$ 
minima
were used (the latter is only for $V^{gg}$).
The third approach can provide coupling parameters with much smaller amount of calculations compared with methods (i) and (ii), nonetheless, the relaxed structures might be local minima. 
Besides the expectation value of the quadratic vibronic operator, the pseudo JT effect may give important contribution to the quadratic coupling parameters \cite{Garcia-Fernandez2005}.
In the present approach, the latter is also included in the obtained quadratic coupling parameters because the first principles APES should contain it. 

The derived Hessian and the vibronic coupling constants are tabulated in Supplemental Materials \cite{SM} (Tables S6-8). 
The order of the quadratic vibronic coupling parameters is at most 10$^{-7}$ a.u., which is about 10-100 times smaller than $K^\Gamma$.
Therefore, as seen from Eqs. (\ref{Eq:UD5d}), (\ref{Eq:UD3d}) and (\ref{Eq:UD2h}), the warping of the APES is small in comparison with the linear JT stabilization energy. 
The influence of the $g_g$ modes on the APES (\ref{Eq:DeltaU}) is very small, nonetheless, there are small contributions to the deformation \cite{SM}.
Although the DFT calculations were carefully performed with fine grid and tight condition for the convergence of self-consistent field calculations (highest accuracy within the code \cite{g16}), the values of the quadratic coupling parameters depend on the approaches (i)-(iii). 
Given the smallness of the obtained parameters, more accurate calculations would be necessary for the derivation of the reliable quadratic vibronic coupling parameters.

\begin{table}[tb]
\caption{
The JT stabilization energies of C$_{60}^-$ with respect to the symmetrized distortions (meV). 
The origin of the energy corresponds to 
that of 
the undeformed structure. 
(i)-(iii) indicate the methods described in Sec. \ref{Sec:method}.
}
\label{Table:SJT}
\begin{ruledtabular}
\begin{tabular}{cccc}
         & (i) & (ii) & (iii) \\
\hline
$D_{5d}$ & $-47.06$ & $-49.38$ & $-48.00$ \\
$D_{3d}$ & $-48.88$ & $-48.44$ & $-48.49$ \\
$D_{2h}$ & $-48.56$ & $-48.58$ & $-48.40$ \\ 
\end{tabular}
\end{ruledtabular}
\end{table}

\section{Quadratic Jahn-Teller effect in C$_{60}^{n-}$ anions}
\subsection{Adiabatic potential energy surface}
\label{Sec:APES_C60-}
With the obtained three sets of vibronic parameters, the warping of the APES of C$_{60}^-$ is analyzed. 
The space for the analysis of 
the
APES's of $T_{1u} \otimes 8h_g$ JT model can be significantly reduced by employing the symmetrized distortions (Appendix \ref{A:q}).
The obtained energies are shown in Table \ref{Table:SJT}. 
In the case of parameters (i) and (iii), the $D_{3d}$ minima is more stable than that of $D_{5d}$ and vice versa for (ii). 
Thus, we cannot conclude about the symmetry of the JT deformation at the global minima as well as the stabilization energy. 
However, the energy difference between these extrema is less than 2 meV, which is much smaller than the JT stabilization energy of ca 50 meV for C$_{60}^-$.

The JT distortion of C$_{60}^-$ has been intensively investigated computationally in the past.
Our result shows that the $D_{5d}$ minima are more stable by 0.31 meV than the $D_{3d}$ minima. 
The result disagrees with some the previous calculations. 
Within semiempirical modified intermediate neglect of differential overlap (MINDO) \cite{Tanaka1992}, 
and generalized gradient approximation (GGA) \cite{Kern2013}, 
the
$D_{3d}$ minima were concluded to be the lowest. 
Within the unrestricted Hartree-Fock (HF) calculations, the $D_{3d}$ and $D_{2h}$ minima have the same energies and the both are lower than the $D_{5d}$ minima by 0.2 meV \cite{Koga1992}. 
Previous B3LYP calculations also predict the $D_{3d}$ as the global minima, albeit not very trustfully because the reported energy gap between the $D_{3d}$ and $D_{5d}$ minima is as much as 40 meV \cite{Saito2002}. 
Our calculations show that the $D_{5d}$ minima 
reported in Ref. \cite{Saito2002} is
not the lowest ones. 
Recent calculations within local density approximation (LDA) approximation 
predict that the $D_{2h}$ minima is more stable than the $D_{3d}$ and $D_{5d}$ minima by 0.5 and 1.5 meV, respectively \cite{Ramanantoanina2013} (The $D_{3d}$ minima is unstable and the global $C_{2h}$ symmetric one is lower by 0.6 meV and exists close to the $D_{3d}$).
However, the order of the $D_{5d}$, $D_{3d}$ and $D_{2h}$ minima disagrees with the one expected for the $T_{1u} \otimes h_g$ Jahn-Teller model \cite{Dunn1995}. 

Finding global minima of C$_{60}^-$ with first principles calculations is not an easy task because of the presence of many local minima in the APES. 
However, taking into account also the previous calculations, $D_{3d}$ minima seems to correspond to the global ones.
Compared to the DFT/HF calculations, our estimate of the warping of 2 meV could be regarded as the upper limit.

\begin{figure}[tb]
\includegraphics[width=8cm]{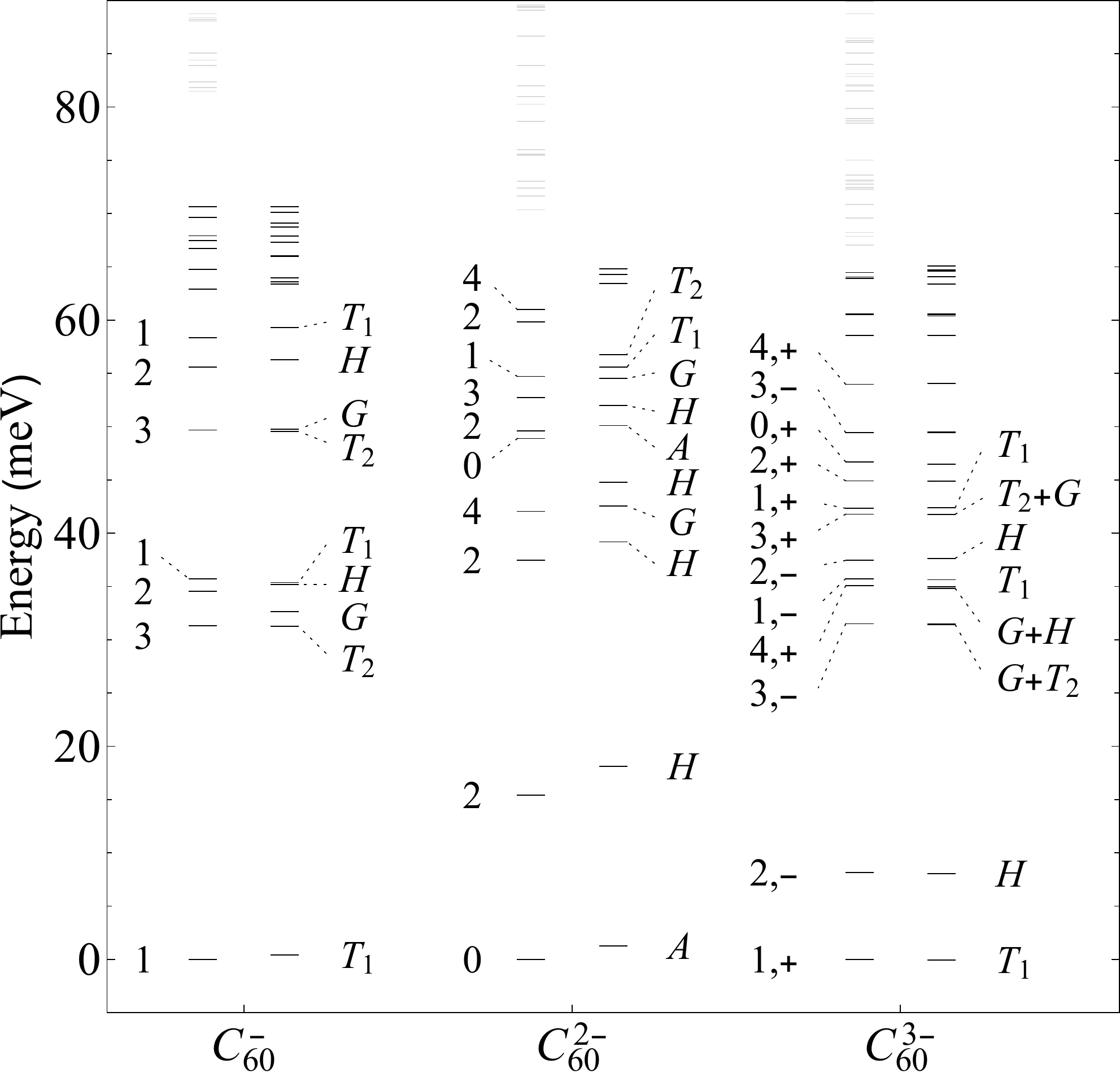}
\caption{
Vibronic levels without and with quadratic vibronic coupling (meV). 
For each anion, the linear (left) and quadratic (right) vibronic levels are shown. 
Among linear vibronic levels, black (light gray) lines show the vibronic levels which are (not) taken into account to calculate the quadratic vibronic levels. 
The quadratic coupling parameters obtained with method (i) are used. 
The vibronic levels are shown with respect to the ground linear vibronic level for each anion. 
}
\label{Fig:E}
\end{figure}

\subsection{Vibronic states}
\label{Sec:Vibro_C60-}
The vibronic states of C$_{60}$ anions were calculated using the DFT-derived coupling parameters. 
Using the linear vibronic states presented in Ref. \cite{Liu2018}, the matrix elements of $\hat{M}^{\Gamma_1 \Gamma_2}_{\nu \mu_1 \mu_2} = \partial \hat{H}^{(2)}_\text{JT}/\partial V^{\Gamma_1 \Gamma_2}_{\nu \mu_1 \mu_2}$ were calculated.
The calculated elements are listed in Tables S10-12 in Supplemental Materials \cite{SM}, and hence, one can calculate the low-lying vibronic states with other sets of quadratic coupling parameters.
Figure \ref{Fig:E} shows the linear vibronic levels from Ref. \cite{Liu2018} and the obtained vibronic levels with parameters' set (i). 
The latter is chosen because it gives the largest change in vibronic levels among 
the
three sets of parameters and also it gives the $D_{3d}$ global minima (for other cases see \cite{SM}).

The vibronic levels of C$_{60}^-$ are shifted at most by 1-2 meV. 
The shift is enhanced in C$_{60}^{2-}$ as expected. 
However, in both cases, the shift of levels is much smaller than the energy gap between the ground and low-lying vibronic levels. 
Besides the weak quadratic vibronic coupling, the vibronic reduction of the matrix elements of $\hat{M}^{\Gamma_1 \Gamma_2}_{\nu \mu_1 \mu_2}$ further quenches the effect of 
the
quadratic coupling. 
In the case of C$_{60}^{3-}$, the quadratic vibronic coupling changes the vibronic energy quadratically [Eq. (\ref{Eq:mat})], resulting in very small shift.

As shown above, in all C$_{60}^{n-}$ anions, the vibronic states are not modified much by the quadratic coupling. 
Thus, except for the molecular spectroscopic studies with very high resolution, the linear vibronic states are sufficient to quantitatively understand the properties of C$_{60}$ anions. 
This also means that the JT dynamics of free C$_{60}$ anions is closer to the pseudorotational type than to the tunneling dynamics type. 
Although the dynamical JT effect is influenced by the environments, in cubic trivalent fullerides the JT dynamics remains of pseudorotational type because of the strong linear JT stabilization and negligible effect of quadratic vibronic coupling.
The higher order vibronic coupling may play important role in the study of high-resolution spectroscopy and may also be enhanced in low-symmetric environment. 
In the former, the quadratic coupling governs the splitting pattern of the highly degenerate excited vibronic levels (Fig. \ref{Fig:E}).
In the latter,
the JT dynamics can be modified in low-symmetric environment such as solution \cite{Hands2008, Kundu2015}, matrix \cite{Kato1993, Kondo1995, Kern2013}, and surface \cite{Wachowiack2005, Dunn2015}.
Quantitative understanding of these situations is beyond the scope of this work.

\section{Conclusion}
\label{Conclusion}
We studied the quadratic JT effect of C$_{60}^{n-}$ anions ($n =$ 1-5) based on DFT calculations.
The main results are the following:
\begin{enumerate}
 \item Analysis of the APES of the $T_{1u} \otimes (g_g \oplus h_g)$ JT model. 
 \item Derivation of the selection rules for the matrix elements of the quadratic vibronic contribution in the basis of the linear vibronic states.
 \item Calculation of the matrix elements of $\hat{H}^{(2)}_\text{JT}$ in the basis of linear vibronic states.
 \item DFT derivation of the quadratic orbital vibronic coupling constants of C$_{60}$ anions.
 \item Evaluation of the effect of the quadratic coupling on the low-lying vibronic levels.
\end{enumerate}
Although the DFT-derived quadratic vibronic coupling parameters are not sufficiently accurate, it was revealed that the quadratic coupling gives much weaker effect than 
the
linear vibronic couplings do in C$_{60}^{n-}$ anions. 
Because of the symmetry and vibronic reduction as well as the weak 
quadratic 
vibronic
couplings, the linear vibronic states of C$_{60}$ anions are not modified much.
In particular, the effect of 
the
quadratic coupling is significantly reduced in C$_{60}^{3-}$, resulting in robust pseudorotational vibronic dynamics in alkali-doped fullerides. 
Besides, the effect of the JT inactive $g_g$ modes on the APES was analyzed. 
The contribution from the $g_g$ mode to the APES is negligible in C$_{60}$ anions, while, it could be more important in icosahedral metal clusters where quadratic coupling is expected to be stronger \cite{Yi1991, Wang2002}.

\section*{Acknowledgment}
D.L. gratefully acknowledges funding by the China Scholarship Council.
N.I. is supported by Japan Society for the Promotion of Science Overseas Research Fellowship.

\appendix

\section{Symmetrized displacement}
\label{A:q}
The JT deformations which keep the symmetry of a subgroup $G$ of $I_h$ are given here ($G = D_{5d}, D_{3d}, D_{2h}$). 
The expression of the symmetrized $h_g$ displacements 
has
been presented in Refs. \cite{Dunn1995, Hands2006, Alqannas2013}.
The following symmetrized displacements contain those by $g_g$ modes as well as the $h_g$ displacements.
The pattern of the deformations 
can be
obtained by using the relation between the spherical harmonics and polynomials: 
Substituting the direction of 
the
$C_5$, $C_3$, $C_2$ axes in the Cartesian coordinate ($t_{1u}$ representation) into the spherical harmonics of 2nd rank, the five components of the spherical harmonics correspond to the patterns of symmetrized $h_g$ deformation. 
The symmetrized $g_g$ deformation is obtained by combining the similar procedure and the relation between the 4th rank spherical harmonics and $G_g \oplus H_g$.
Below, we choose the deformation which keeps the symmetry axes in the $zx$ plane, and the $a_g$ contributions are not explicitly written because they do not lower the symmetry.

Under $I_h \downarrow D_{5d}$, the $g_g$ and $h_g$ representations are reduced as \cite{Altmann1994}
\begin{eqnarray}
 g_g\downarrow D_{5d} &=& e_{1g} \oplus e_{2g},
\nonumber\\
 h_g\downarrow D_{5d} &=& a_g \oplus e_{1g} \oplus e_{2g}.
\end{eqnarray}
Thus, the $D_{5d}$ distortion can be expressed by using one coordinate for each $h_g(\mu)$ mode:
\begin{eqnarray}
 \Delta \bm{R}_{D_{5d}} &=& \frac{1}{\sqrt{M}} \sum_{\mu = 1}^8 q_{h_g(\mu)}
 \left[ \frac{\phi^2}{2\sqrt{5}} \bm{e}_{h_g(\mu)\theta} 
 \right. 
\nonumber\\
 &+& 
 \left.
 \frac{\phi^{-1}}{2} \sqrt{\frac{3}{5}} \bm{e}_{h_g(\mu)\epsilon}
+ \sqrt{\frac{3}{5}} \bm{e}_{h_g(\mu)\eta} \right].
\label{Eq:RD5d}
\end{eqnarray}
Here, $M$ is the mass of carbon atom. 
On the other hand, under $I_h \downarrow D_{3d}$, both 
the
$g_g$ and $h_g$ representations contain the totally symmetric representation \cite{Altmann1994}:
\begin{eqnarray}
 g_g\downarrow D_{3d} &=& a_{1g} \oplus a_{2g} \oplus e_g,
\nonumber\\
 h_g\downarrow D_{3d} &=& a_{1g} \oplus 2e_g.
\end{eqnarray}
Therefore, the $D_{3d}$ deformation is expressed by one $g_g$ and one $h_g$ coordinates \cite{Altmann1994}: 
\begin{eqnarray}
 \Delta \bm{R}_{D_{3d}} &=& \frac{1}{\sqrt{M}} \sum_{\mu = 1}^8 q_{h_g(\mu)}
 \left[ -\frac{\phi^{-1}}{2} \bm{e}_{h_g(\mu)\theta} 
\right.
\nonumber\\
 &+& 
\left.
 \frac{\phi^2}{2\sqrt{3}} \bm{e}_{h_g(\mu)\epsilon} 
+ \frac{1}{\sqrt{3}} \bm{e}_{h_g(\mu)\eta} \right]
\nonumber\\
 &+&
  \frac{1}{\sqrt{M}} \sum_{\mu = 1}^6 q_{g_g(\mu)}
 \left[ \frac{1}{\sqrt{6}} \bm{e}_{g_g(\mu)a} - \sqrt{\frac{5}{6}} \bm{e}_{g_g(\mu)y}\right].
\nonumber\\
\label{Eq:RD3d}
\end{eqnarray}
Finally, under $I_h \downarrow D_{2h}$, 
\begin{eqnarray}
 g_g\downarrow D_{2h} &=& a_g \oplus b_{1g} \oplus b_{2g} \oplus b_{3g}, 
\nonumber\\
 h_g\downarrow D_{2h} &=& 2a_g \oplus b_{1g} \oplus b_{2g} \oplus b_{3g}.
\end{eqnarray}
Therefore, the $g_g$ and $h_g$ deformations are expressed by one ($q_{g_ga}$) and two ($q_{h_g \theta}, q_{h_g\epsilon}$) coordinates, respectively:
\begin{eqnarray}
 \Delta \bm{R}_{D_{2h}} &=& \frac{1}{\sqrt{M}} \sum_{\mu = 1}^8 \left[ q_{h_g(\mu)\theta} \bm{e}_{h_g(\mu)\theta} + q_{h_g(\mu)\epsilon} \bm{e}_{h_g(\mu)\epsilon} \right]
\nonumber\\
 &+& \frac{1}{\sqrt{M}} \sum_{\mu = 1}^6 q_{g_g(\mu)a} \bm{e}_{g_g(\mu)a}.
\label{Eq:RD2h}
\end{eqnarray}
In the above equations, $\Delta \bm{R}_G$ is the 180 dimensional vector consisting of all Cartesian coordinates of carbon atoms in C$_{60}$, $\bm{e}_{\Gamma(\mu)\gamma}$ is the polarization vector, $q_{\Gamma(\mu)}$ is the magnitude of the displacement, and $\phi = (1+\sqrt{5})/2$.

\section{Derivatives of the APES}
\label{A:Hessian}
The second 
derivatives
of the APES with respect to $a_g$, $g_g$, and $h_g$ coordinates are given.
Below, the APES around the minima with symmetry $G$ is written as $U_G$ ($G = D_{5d}, D_{3d}, D_{2h}$).
The notation of the coordinates are the same as in Eqs. (\ref{Eq:RD5d}), (\ref{Eq:RD3d}), and (\ref{Eq:RD2h}).
At the $D_{5d}$ structure, 
\begin{eqnarray}
 \frac{\partial^2 U_{D_{5d}}}{\partial q_{a_g(\mu_1)} \partial q_{a_g(\mu_2)}} &=& K^{a_g}_{\mu_1 \mu_2}, 
\nonumber\\
 \frac{\partial^2 U_{D_{5d}}}{\partial q_{a_g(\mu_1)} \partial q_{h_g(\mu_2)}} &=& \frac{1}{2} V^{a_g h_g}_{\mu_1 \mu_2},
\nonumber\\
 \frac{\partial^2 U_{D_{5d}}}{\partial q_{h_g(\mu_1)} \partial q_{h_g(\mu_2)}} &=& K^{h_g}_{\mu_1 \mu_2} + \frac{2}{\sqrt{5}} V^{h_g h_g}_{1\mu_1 \mu_2}.
\end{eqnarray}
At the $D_{3d}$ structure, 
\begin{eqnarray}
 \frac{\partial^2 U_{D_{3d}}}{\partial q_{a_g(\mu_1)} \partial q_{a_g(\mu_2)}} &=& K^{a_g}_{\mu_1 \mu_2}, 
\nonumber\\
 \frac{\partial^2 U_{D_{3d}}}{\partial q_{a_g(\mu_1)} \partial q_{h_g(\mu_2)}} &=& -V^{a_g h_g}_{\mu_1 \mu_2},
\nonumber\\
 \frac{\partial^2 U_{D_{3d}}}{\partial q_{g_g(\mu_1)} \partial q_{g_g(\mu_2)}} &=& K^{g_g}_{\mu_1\mu_2} + \frac{5}{3} V^{g_g g_g}_{\mu_1 \mu_2},
\nonumber\\
 \frac{\partial^2 U_{D_{3d}}}{\partial q_{g_g(\mu_1)} \partial q_{h_g(\mu_2)}} &=& -\frac{2}{3} V^{g_g h_g}_{1\mu_1 \mu_2},
\nonumber\\
 \frac{\partial^2 U_{D_{3d}}}{\partial q_{h_g(\mu_1)} \partial q_{h_g(\mu_2)}} &=& K^{h_g}_{\mu_1\mu_2} - \frac{4}{3} V^{h_g h_g}_{2\mu_1 \mu_2}.
\end{eqnarray}
At the $D_{2h}$ structure, 
\begin{eqnarray}
 \frac{\partial^2 U_{D_{2h}}}{\partial q_{a_g(\mu_1)} \partial q_{a_g(\mu_2)}} &=& K^{a_g}_{\mu_1 \mu_2}, 
\nonumber\\
 \frac{\partial^2 U_{D_{2h}}}{\partial q_{a_g(\mu_1)} \partial q_{h_g(\mu_2)}} &=& -V^{a_g h_g}_{\mu_1 \mu_2}, 
\nonumber\\
 \frac{\partial^2 U_{D_{2h}}}{\partial q_{g_g(\mu_1)a} \partial q_{g_g(\mu_2)a}} &=& K^{g_g}_{\mu_1 \mu_2}, 
\nonumber\\
 \frac{\partial^2 U_{D_{2h}}}{\partial q_{g_g(\mu_1)a} \partial q_{h_g(\mu_2)\theta}} &=& -\frac{1}{2}\sqrt{\frac{3}{2}} V^{g_g h_g}_{1 \mu_1 \mu_2}, 
\nonumber\\
 \frac{\partial^2 U_{D_{2h}}}{\partial q_{g_g(\mu_1)a} \partial q_{h_g(\mu_2)\epsilon}} &=& -\frac{1}{2}\sqrt{\frac{5}{2}} V^{g_g h_g}_{2 \mu_1 \mu_2}, 
\nonumber\\
 \frac{\partial^2 U_{D_{2h}}}{\partial q_{h_g(\mu_1)\theta} \partial q_{h_g(\mu_2)\theta}} &=& K^{h_g}_{\mu_1 \mu_2} - \frac{\sqrt{5}}{8} V^{h_g h_g}_{1 \mu_1 \mu_2} 
- \frac{9}{8} V^{h_g h_g}_{2 \mu_1 \mu_2},
\nonumber\\
 \frac{\partial^2 U_{D_{2h}}}{\partial q_{h_g(\mu_1)\theta} \partial q_{h_g(\mu_2)\epsilon}} &=& - \frac{3}{8}\sqrt{3} V^{h_g h_g}_{1 \mu_1 \mu_2} 
+ \frac{\sqrt{15}}{8} V^{h_g h_g}_{2 \mu_1 \mu_2},
\nonumber\\
 \frac{\partial^2 U_{D_{2h}}}{\partial q_{h_g(\mu_1)\epsilon} \partial q_{h_g(\mu_2)\epsilon}} &=& K^{h_g}_{\mu_1 \mu_2} + \frac{\sqrt{5}}{8} V^{h_g h_g}_{1 \mu_1 \mu_2} 
+ \frac{9}{8} V^{h_g h_g}_{2 \mu_1 \mu_2}.
\nonumber\\
\end{eqnarray}


%

\end{document}